\documentclass{aa}

\usepackage{graphicx}
\usepackage{txfonts}
\usepackage{color}
\usepackage{kotex}   

\begin{document}

   \title{Search for Exoplanets around Northern Circumpolar Stars}

   \subtitle{V. Three likely planetary companions to the giant stars HD 19615, HD 150010, and HD~174205}

   \author{G. Jeong (정광희)\inst{1},
          B.$-$C. Lee (이병철)\inst{2,3},
          M.$-$G. Park (박명구)\inst{4},
          T.$-$Y. Bang (방태양)\inst{4},
          \and
          I. Han (한인우)\inst{2}
          }

   \institute{Antbridge Technology, 79-1, 101 Gajeong-ro Yuseong-gu, Daejeon 34120, Korea\\
              \email{tlotv1@gmail.com}
            \and
              Korea Astronomy and Space Science Institute, 776,
                Daedeokdae-Ro, Youseong-Gu, Daejeon 34055, Korea\\
              \email{[bclee;iwhan]@kasi.re.kr}
            \and
                Korea University of Science and Technology, Gajeong-ro Yuseong-gu, Daejeon 34113, Korea
            \and
             Department of Astronomy and Atmospheric Sciences, Kyungpook National University, Daegu 41566, Korea\\
              \email{mgp@knu.ac.kr;apollo.choe@gmail.com}
             }

   \date{Received October 6, 2021; accepted November 9, 2021}


  \abstract
   {}
   {We report the detection of long-period radial velocity (RV) variations in three giant stars, HD~19615, HD~150010, and HD~174205, using precise RV measurements.}
   {
   These detections are part of the Search for Exoplanets around Northern Circumpolar Stars (SENS) survey being conducted at the Bohyunsan Optical Astronomy Observatory (BOAO).
   The nature of the RV variations was investigated by analyzing the photometric and line shape variations.
   We found no variability with the RV period in these quantities and conclude that the RV variations are most likely caused by planetary companions.
}
   {Orbital solutions for the three stars yield orbital periods of 402 d, 562 d, and 582 d and minimum masses of 8.5 $M_{J}$, 2.4 $M_{J}$, and 4.2 $M_{J}$, respectively. These masses and periods are typical for planets around intermediate-mass stars, although some unclear interpretations and recent studies may being calling some planet convictions  into question.\ Nevertheless, the SENS program is contributing to our knowledge of giant planets around intermediate-mass stars.}
   {}
   \authorrunning{Jeong et al.}
   \titlerunning{Search for Exoplanets around Northern Circumpolar Stars}
   \maketitle
%

\section{Introduction}

The measurements of stellar radial velocity (RV) are one of the most successful techniques employed in  the search for exoplanets.
We have been conducting an exoplanet search program around late-type giant stars since 2004.
This program has made substantial contributions to both exoplanet and asteroseismic studies around giant stars
\citep{2010A&A...509A..24H,2011A&A...529A.134L,2012A&A...546A...5L,2012A&A...548A.118L,2013A&A...549A...2L,2014A&A...566A..67L,2014JKAS...47...69L}.
However, these works were occasionally hampered by an aliasing effect caused by the seasonal dependence of observations. To cope with this bias, we started the Search for Exoplanets around Northern Circumpolar Stars (SENS; \citealt{2015A&A...584A..79L}) survey in 2010. Because all targets of the SENS program can be observed during all seasons throughout the year, the seasonal bias is minimized.  Furthermore, this strategy improved the observation efficiency by reducing the telescope slew time. (\citealt{2012A&A...546A...5L})

Here we present the detection of three new planet candidates. Section 2 introduces the observation strategy. The stellar properties and the analyses are described in detail in Sect. 3. We report the analyses of RV variations and the results of the measurements in Sect. 4, and, finally, our discussion of the results is presented in Sect. 5.

\section{Observations}

\begin{table*}
\renewcommand{\thetable}{\arabic{table}}
\centering
\caption{RV measurements for HD 19615 from February 2010 to March 2021 using the BOES.} \label{tab:rv1}
\begin{tabular}{ccccccccc}
\hline
JD & RV  & $\pm \sigma$ & JD & RV  & $\pm \sigma$  & JD & RV  & $\pm \sigma$\\
$-$2,450,000 &{m\,s$^{-1}$}& {m\,s$^{-1}$}& $-$2,450,000 & {m\,s$^{-1}$} & {m\,s$^{-1}$} & $-$2,450,000 & {m\,s$^{-1}$} & {m\,s$^{-1}$}\\
\hline
5250.122138  &    172.1  &     13.9  &   7327.282286  &    375.8  &     11.7 &   8932.016082 &     297.0   &    18.9   \\
5842.239861  &   -135.9  &     10.4  &   7704.032537  &    419.6  &     12.1 &   8933.033641 &     331.2   &    26.2   \\
5933.110447  &   -269.3  &     13.1  &   7757.119503  &     90.8  &     10.8 &   8942.963325 &     198.6   &    15.7   \\
5962.989966  &   -192.3  &     12.3  &   7758.088199  &     34.1  &     13.8 &   8945.014470 &     201.6   &    17.1   \\
6259.132566  &   -256.3  &     10.0  &   7813.013370  &    -62.4  &     11.5 &   8969.961480 &     178.8   &    19.5   \\
6287.044466  &   -293.7  &     18.5  &   7855.011677  &    -81.4  &     22.9 &   9149.317839 &    -168.9   &    12.2   \\
6288.161790  &   -166.4  &      9.5  &   7855.986123  &   -155.7  &     15.2 &   9150.041118 &    -190.8   &    21.3   \\
6347.047078  &   -101.5  &     12.9  &   8015.212962  &     48.9  &     11.8 &   9151.008641 &    -184.4   &    11.1   \\
6551.243387  &    133.2  &     11.9  &   8092.145036  &    174.8  &     13.6 &   9161.097381 &    -177.0   &    12.6   \\
6578.277410  &    -13.3  &     14.4  &   8109.225577  &     90.8  &     12.8 &   9161.108724 &    -148.6   &    12.8   \\
6617.007990  &    -98.8  &      9.5  &   8516.945624  &     88.1  &     12.0 &   9161.097381 &    -177.0   &    12.6   \\
6714.072126  &   -109.5  &     17.3  &   8561.987563  &    -13.2  &     15.0 &   9161.108724 &    -148.6   &    12.8   \\
6739.961457  &   -120.5  &     10.6  &   8830.238542  &    -78.7  &     17.6 &   9162.194690 &    -141.3   &    13.0   \\
6808.255282  &   -147.6  &     20.5  &   8849.167430  &    121.2  &     41.3 &   9217.071292 &    -132.7   &    15.4   \\
6922.137214  &    106.6  &     11.8  &   8852.045193  &     49.0  &     14.2 &   9218.071261 &    -167.9   &    15.0   \\
6965.371019  &    234.8  &     12.4  &   8862.029439  &    117.6  &     13.6 &   9298.983663 &      30.0   &    15.2   \\
7094.012094  &   -227.1  &     12.9  &   8863.129164  &    161.4  &     16.5 &               &             &           \\
7298.026234  &    410.2  &     10.7  &   8893.969101  &    201.4  &     16.5 &               &             &           \\

\hline
\end{tabular}
\end{table*}

\begin{table*}
\renewcommand{\thetable}{\arabic{table}}
\centering
\caption{RV measurements for HD 150010 from June 2010 to June 2021 using the BOES.} \label{tab:rv2}
\begin{tabular}{ccccccccc}
\hline
JD & RV  & $\pm \sigma$ & JD & RV  & $\pm \sigma$  & JD & RV  & $\pm \sigma$\\
$-$2,450,000 &{m\,s$^{-1}$}& {m\,s$^{-1}$}& $-$2,450,000 & {m\,s$^{-1}$} & {m\,s$^{-1}$} & $-$2,450,000 & {m\,s$^{-1}$} & {m\,s$^{-1}$}\\
\hline

5357.027343  &     11.3  &     18.6   &   7475.184243  &    -14.8  &     17.4 &    8943.045719  &     35.5  &     16.9  \\
5616.392269  &    143.6  &     25.1   &   7527.109889  &    -40.3  &     17.5 &    8943.066656  &     25.6  &     15.9  \\
5671.011796  &    163.2  &     24.4   &   7530.213694  &    -19.7  &     19.3 &    8944.157323  &     54.2  &     19.0  \\
5842.957103  &      4.6  &     20.8   &   7704.941274  &     19.5  &     14.1 &    8945.081580  &     77.2  &     16.9  \\
6379.262625  &     51.6  &     14.6   &   7819.259376  &      9.2  &     15.0 &    8969.097550  &     23.8  &     18.2  \\
6409.327670  &     -1.2  &     14.6   &   7856.248364  &    -18.3  &     15.2 &    8970.135878  &     18.2  &     17.5  \\
6583.040523  &     60.6  &     25.3   &   7895.087187  &    -46.2  &     16.7 &    9008.990863  &     77.3  &     22.6  \\
6710.321857  &     99.2  &     15.0   &   8010.998230  &    -58.6  &     13.2 &    9128.930702  &     -2.4  &     16.6  \\
6739.165553  &     46.3  &     15.1   &   8037.017169  &    -84.1  &     18.4 &    9148.968556  &    -37.8  &     12.4  \\
6805.184754  &     42.4  &     18.1   &   8105.359985  &    -47.5  &     15.2 &    9149.918236  &    -51.0  &     14.6  \\
6921.965641  &   -102.7  &     16.2   &   8743.003171  &   -103.8  &     14.7 &    9150.910347  &    -53.3  &     13.5  \\
6963.939995  &    -36.9  &     19.6   &   8772.957082  &    -32.9  &     13.7 &    9160.945100  &    -74.0  &     19.0  \\
6970.901351  &    -80.4  &     15.8   &   8852.378263  &     11.7  &     17.2 &    9161.952258  &    -54.6  &     15.9  \\
7027.383448  &    -73.2  &     16.5   &   8863.303083  &    -49.7  &     17.0 &    9296.119476  &    -75.0  &     20.6  \\
7067.324662  &    -20.2  &     20.7   &   8898.199527  &      6.9  &     21.4 &    9330.007068  &    -81.1  &     61.1  \\
7148.089384  &     -8.5  &     17.7   &   8927.267737  &      4.0  &     20.0 &    9359.173258  &     10.2  &     20.4  \\
7169.267191  &     86.2  &     18.3   &   8932.072411  &     56.9  &     22.5 &    9370.048097  &     51.5  &     44.1  \\
7298.972916  &      8.8  &     15.2   &   8933.077617  &     80.6  &     23.7 &    9370.173096  &    -13.0  &     16.5  \\

\hline
\end{tabular}
\end{table*}

\begin{table*}
\renewcommand{\thetable}{\arabic{table}}
\centering
\caption{RV measurements for HD 174205 from February 2010 to June 2021 using the BOES.} \label{tab:rv3}
\begin{tabular}{ccccccccc}
\hline
JD & RV  & $\pm \sigma$ & JD & RV  & $\pm \sigma$  & JD & RV  & $\pm \sigma$\\
$-$2,450,000 &{m\,s$^{-1}$}& {m\,s$^{-1}$}& $-$2,450,000 & {m\,s$^{-1}$} & {m\,s$^{-1}$} & $-$2,450,000 & {m\,s$^{-1}$} & {m\,s$^{-1}$}\\
\hline

5248.349219  &     76.0   &     9.3 &    7703.935737  &    33.9   &     8.2  &     8933.159881   &   103.4   &    12.1     \\
5456.162247  &     13.0   &     7.4 &   7705.013445   &    44.3   &     7.7  &     8943.148932   &    31.1   &     9.5     \\
5672.200478  &    -67.8   &    11.4 &   7756.929401   &    42.4   &     8.7  &     8943.164175   &    39.1   &     9.5     \\
5844.052953  &     19.8   &     9.0 &   7757.942321   &    60.4   &    11.8  &     8944.179621   &    84.1   &    12.5     \\
6287.934421  &    -71.1   &     8.7 &   7856.322146   &    15.5   &     9.6  &     8944.194933   &    92.4   &    10.4     \\
6410.265546  &      7.5   &     8.5 &   7893.146122   &   -25.4   &     8.9  &     8945.143700   &    69.8   &     9.0     \\
6681.350238  &     36.5   &    10.6 &   7934.189754   &    -4.9   &     9.0  &     8969.130023   &    41.5   &    11.4     \\
6739.189411  &     59.9   &    11.8 &   8015.076632   &   -99.0   &     7.8  &     8970.154940   &    96.6   &    11.2     \\
6805.230356  &     24.3   &    10.6 &   8091.946858   &   -60.9   &    11.5  &     9128.947432   &   -42.3   &     9.1     \\
6921.989868  &    -45.8   &     8.7 &   8109.912914   &   -11.3   &     9.0  &     9148.987572   &   -29.7   &    10.7     \\
6967.945510  &    -88.8   &     8.0 &   8743.048983   &    31.0   &     7.6  &     9149.969785   &   -33.4   &     8.5     \\
7066.310819  &    -93.7   &    14.4 &   8772.993155   &    27.0   &     8.7  &     9150.957381   &    -8.6   &     9.8     \\
7068.260174  &    -61.8   &    10.1 &   8826.877616   &    24.2   &     9.2  &     9152.007233   &   -90.8   &     8.8     \\
7148.147847  &      4.3   &    10.9 &   8829.916120   &    24.1   &     9.6  &     9160.973048   &   -66.7   &    10.3     \\
7171.236702  &     32.3   &     9.5 &   8831.904548   &    33.7   &    11.2  &     9160.987955   &   -75.4   &     8.9     \\
7299.018323  &      0.6   &    11.7 &   8852.388466   &   -16.7   &    11.9  &     9161.975479   &     3.9   &    10.7     \\
7475.271895  &   -139.2   &    10.7 &   8863.322980   &    48.8   &    14.3  &     9296.267940   &   -24.6   &    11.1     \\
7525.170621  &   -125.0   &     7.2 &   8898.315208   &    46.5   &    13.1  &     9359.214704   &     6.6   &    12.6     \\
7530.266309  &    -94.9   &     8.6 &   8933.146224   &    73.4   &    13.5  &     9370.212997   &    29.1   &     9.0     \\

\hline
\end{tabular}
\end{table*}

Since January 2010, we have collected a total of 52, 54, and 57 spectral data points for \mbox{HD 19615}, \mbox{HD 150010}, and \mbox{HD 174205}, respectively.
The three stars were observed with the high-resolution fiber-fed Bohyunsan Observatory Echelle Spectrograph (BOES; \citealt{2007PASP..119.1052K}) of the 1.8 m telescope at Bohyunsan Optical Astronomy Observatory (BOAO).
Because the candidates are concentrated around the pole star, observations are possible throughout the year. Thus, the SENS survey has an advantage in terms of period coverage.
For precise RV measurements, an iodine absorption (I$_{2}$) cell was placed in the optical path in front of the fiber entrance \citep{2007PKAS...22...75H}.
This is based on the method by \cite{1995PASP..107..966V} and \cite{1996PASP..108..500B}. We also used the matrix formula described by Endl et al. (2000) for the modeling of the instrument profile. And using singular value decomposition instead of the maximum entropy method adopted by \cite{2000A&A...362..585E}, we solved the matrix equation.

The target stars have a V-magnitude range of 6.4--6.7, and we adopted a 200 $\mu$m fiber, an adequate trade-off between the optical efficiency and fiber resolution, which provides a spectral resolution of 45,000.
The typical exposure time is limited to less than 20 minutes, which yields a signal-to-noise ratio of about 150.
In order to check the instrumental stability, we have monitored an RV standard star, $\tau$ Ceti, since 2003. The long-term RV stability of the BOES was $\sim$7 m s$^{-1}$ \citep{2013A&A...549A...2L}.
The RV measurements are listed in Tables~\ref{tab:rv1}-\ref{tab:rv3}.

\section{Stellar properties}
\begin{table*}
\begin{center}
\caption{Stellar parameters for the stars analyzed in the present paper.} \label{tab:ste}
\label{tab5}
\begin{tabular}{lcccc}

\hline
        Parameter       & HD 19615      & HD 150010     & HD 174205             & Ref.\\

\hline
Spectral type                                   & K0            & K2 III        & K2              & 1\\
$\textit{$m_{v}$}$ (mag)                        & 6.697 $\pm$ 0.001             & 6.427 $\pm$ 0.001       & 6.444 $\pm$ 0.001     &  1\\
$\textit{B-V}$ (mag)                            & 1.465 $\pm$ 0.008     & 1.291 $\pm$ 0.008       & 1.234 $\pm$ 0.007     &  1\\
RV (km s$^{-1}$)        & $-$ 35.21 $\pm$ 0.20          & $-$ 35.81 $\pm$ 0.21    & $-$ 5.41 $\pm$ 0.20           &  1\\
$\pi$ (mas)                                     & 3.74 $\pm$ 0.03               & 6.89 $\pm$ 0.03 & 4.24 $\pm$ 0.03       &  2\\
$T_{\rm{eff}}$ (K)              & 3987 $\pm$ 125                & 4256 $\pm$ 125     & 4393 $\pm$ 125        &   2\\
                                                & 4263 $\pm$ 38         & 4153 $\pm$ 83   & 4308 $\pm$ 25 &  3 \\
$\rm{[Fe/H]}$           & $-$ 0.31 $\pm$ 0.11   & $-$ 0.25 $\pm$ 0.22   & $-$ 0.28 $\pm$ 0.33             & 3\\
log $\it g$ (cgs)                       & 1.4 $\pm$ 0.1         & 2.1 $\pm$ 0.1     & 1.8 $\pm$ 0.1 & 3\\
$v_{\rm{micro}}$ (km s$^{-1}$)  & 1.85 $\pm$ 0.12               & 1.27 $\pm$ 0.10    & 1.49 $\pm$ 0.08 & 3\\
Age (Gyr)                                       & 6.7 $\pm$ 1.8         & 4.9 $\pm$ 1.1   & 1.7 $\pm$ 0.3 &  3\\
$\textit{$R_{\star}$}$ ($R_{\odot}$)    & 32.3 $\pm$ 0.9        &  16.2 $\pm$ 0.4     & 26.6 $\pm$ 0.6        & 3\\
$\textit{$M_{\star}$}$ ($M_{\odot}$)    & 1.1 $\pm$ 0.1         & 1.3  $\pm$ 0.1     & 1.8 $\pm$ 0.1 &  3\\
$\textit{$L_{\star}$}$ ($L_{\odot}$)    & 327$\pm$3             & 96$\pm$1      & 239$\pm$2       & 2 \\

$v_{\rm{rot}}$ sin $i$ (km s$^{-1}$)    & 2.6 $\pm$ 0.5         & 2.2 $\pm$ 0.5     & 1.6 $\pm$ 0.5 &  3\\
$P_{\rm{rot}}$ / sin $i$ (days)         & 560   & 391   & 866   &  3\\
\hline
\end{tabular}
\\
\textbf{References.}--- (1) \citet{2012AstL...38..331A}; (2) \citet{2018A&A...616A...1G}; (3) This work.\\

\end{center}
\end{table*}

\begin{table*}
\renewcommand{\thetable}{\arabic{table}}
\centering
\caption{Orbital parameters for the companions.} \label{tab:orb}
\begin{tabular}{lccc}

\hline
    Parameter&  HD 19615 b          &  HD 150010 b     &  HD 174205 b     \\
\hline

P (days)          &    402 $\pm$ 2         & 562 $\pm$ 8     & 582 $\pm$ 17         \\
K (m s$^{-1}$)&   224 $\pm$ 18         & 52 $\pm$ 8      & 76 $\pm$ 11        \\
$T_{periastron}$ (JD) & 2449662 $\pm$ 91 & 2449811 $\pm$175     & 2449918 $\pm$197\\
$e$                      & 0.2 $\pm$ 0.1            & 0.2 $\pm$ 0.1    & 0.4 $\pm$ 0.2       \\
$\omega$ (deg)  & 279 $\pm$ 71           & 137 $\pm$ 91    & 94 $\pm$ 93             \\
$m$ sin $i$ ($M_{J}$) &8.5 $\pm$ 0.7    & 2.4 $\pm$ 0.4         & 4.2 $\pm$ 0.5      \\
$a$ (AU)        & 1.1 $\pm$ 0.1        & 1.4 $\pm$ 0.1      & 1.7 $\pm$ 0.1          \\
$Slope$  (m s$^{-1} yr^{-1}$)& $-$9 $\pm$ 10 & $-$20 $\pm$ 5  & 2 $\pm$ 4      \\
$N_{obs}$           & 52                & 54    & 57     \\
RMS (m s$^{-1}$)        & 85.2          & 36.9  & 39     \\
\hline

\end{tabular}
\end{table*}

We adopted the basic stellar parameters (spectral type, visual magnitude, B-V color index, and RV) from the \emph{HIPPARCOS} catalog \citep{1997yCat.1239....0E} and \citet{2012AstL...38..331A}.
To use more recent effective temperature, parallax, and luminosity data, photometry parameters from Gaia Data Release 2 \citep[DR2;][]{2018A&A...616A...1G} were obtained.

We determined the effective temperature and surface gravity with the measured equivalent widths of \mbox{Fe I} and \mbox{Fe II} lines of each stellar spectrum \citep{2002PASJ...54.1041T}.
The surface gravity was determined from the ionization equilibrium of \mbox{Fe I} and \mbox{Fe II}.
The metallicity ([Fe/H]) and microturbulent velocity ($v_{\rm{micro}}$), including $T_{\rm{eff}}$ and log \textit{g}, were derived from the TGVIT code \citep{2005PASJ...57...27T}.

The stellar age, radius, and mass for each star were calculated using the online tool PARAM 1.3 \footnote{\url{http://stev.oapd.inaf.it/cgi-bin/param_1.3/}} (\citealt{2006A&A...458..609D}) for given parameters $T_{\rm{eff}}$, [Fe/H], $m_{v}$, and $\pi$.
This tool is based on a library of theoretical stellar isochrones (\citealp{2000A&AS..141..371G,2012MNRAS.427..127B}) and the Bayesian estimation method \citep{2005A&A...436..127J}.

Generally, in evolved stars, the low-amplitude, long-period RV variations can originate from the rotational modulation of the surface with inhomogeneities.
We estimated the stellar rotational velocities using the line broadening technique (\citealt{2008PASJ...60..781T}).
The stellar parameters of the observed stars are summarized in Table~\ref{tab:ste}.

\section{Orbital solutions}
We obtained three targets spanning 11 years, which show clear periodic variations.
Initial values for the period were determined via Lomb-Scargle (L-S) analysis, which is appropriate for long-period variations with long-term sporadic observations \citep{1976Ap&SS..39..447L,1982ApJ...263..835S}.
From this initial period, we determined all the orbital elements via fitting with Keplerian orbit models in the \emph{Systemic Console 2} \citep{2009PASP..121.1016M}.
We applied the same method to the three  stars to find the best-fit Keplerian orbital motion for the observed RV variations.
We estimated the parameter uncertainties using the bootstrap routine within \emph{Systemic Console 2} with 10,000 synthetic data set realizations.
The orbital parameters are listed in Table~\ref{tab:orb}.

\begin{figure*}
\includegraphics[width=17cm]{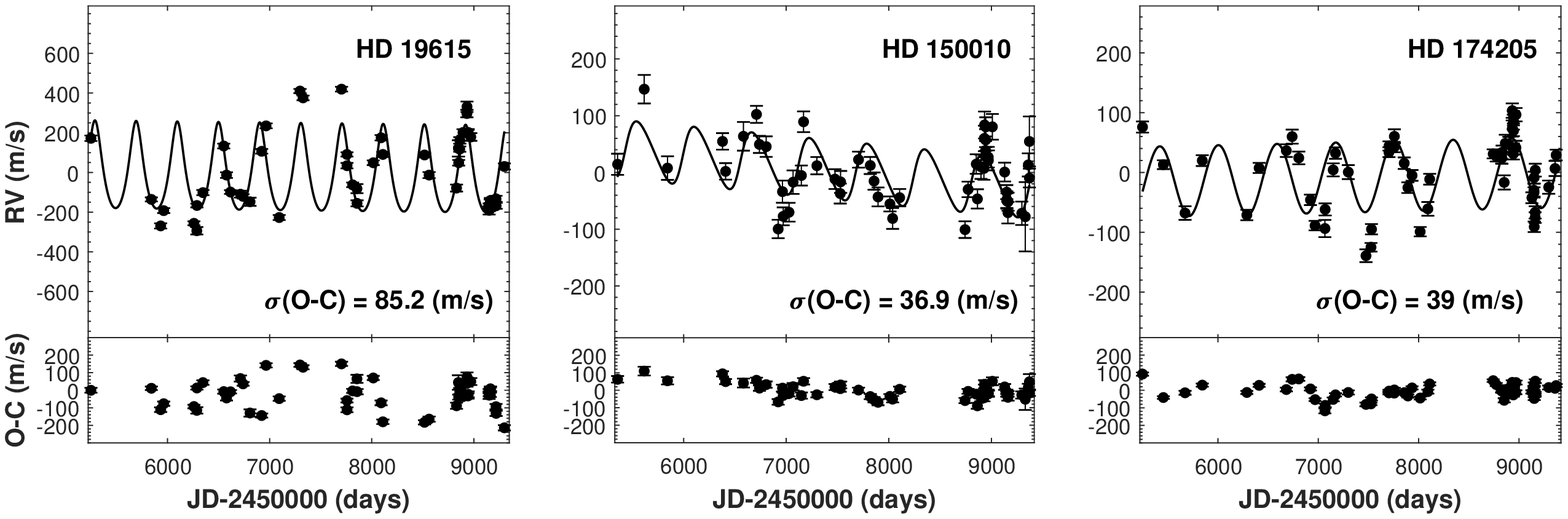}
\center
\caption{Keplerian orbital fits (\emph{top}) and the RV residuals after subtracting the Keplerian fit (\emph{bottom}) for the RVs of the three stars.}
\label{fig:rv_all}
\end{figure*}

\begin{figure*}
\includegraphics[width=17cm]{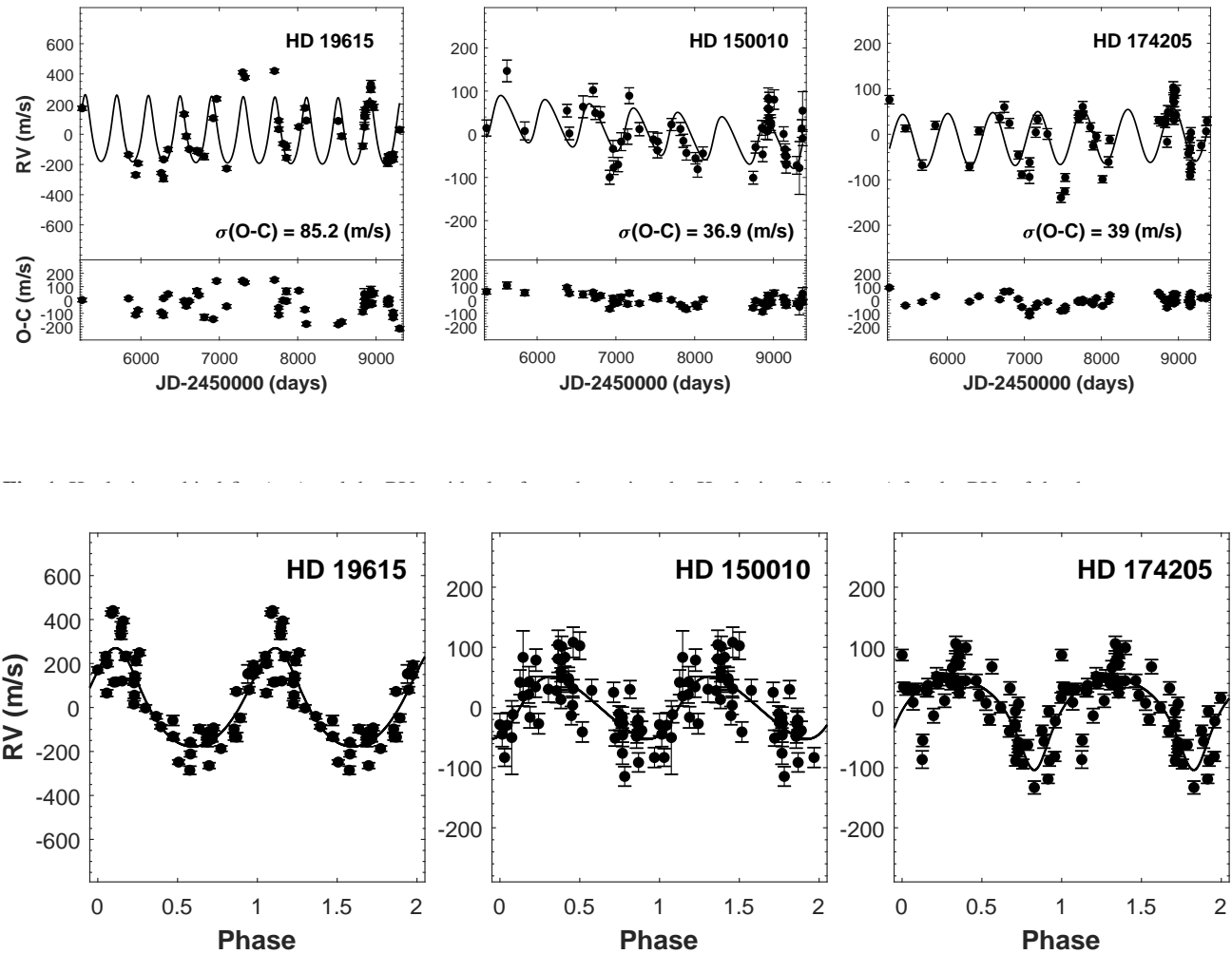}
\caption{Phase-folded RV curves of the three stars.}
\label{fig:pha}
\end{figure*}

The RV data of the three stars are plotted over six cycles along with the best Keplerian orbital fit in the upper panels of Fig.~\ref{fig:rv_all}.
As can be seen from the bottom panel of Fig.~\ref{fig:rv_all}, most of the rms of the RV residuals are significantly larger than the long-term uncertainty measured by the observation of the RV standard star, $\sim$ 7 m s$^{-1}$, and the typical internal error of individual RV precisions, $\sim$ 13 m s$^{-1}$, and they are consistent  with the expected RV scatter (so-called jitter) in giants. Yet they do not show any systemic pattern. These residuals are probably from typical intrinsic RV variations in K giant stars (\citealt{2005PASJ...57...97S,2006A&A...454..943H}).
Furthermore, all of the target stars show a slight linear trend, which may have been caused by an unseen distant companion.
The long-term linear trend is indeed significant and could be caused by a distant companion. \cite{2012AJ....144..165H} reported a distant binary companion to HD 150010 with speckle interferometry, at a projected separation of ~1 arcsecond. This companion may possibly be the source of the long-term linear RV trend.
Figure~\ref{fig:pha} shows the RV measurements phased to the orbital periods for all stars.

\section{Stellar activity and pulsation diagnostics} \label{sec:diag}

The periodic variations of RVs can also be caused by some intrinsic activities of the stars, such as chromospheric activity, stellar pulsation, and rotational modulation of the surface feature. In order to identify the nature of the RV variations, we investigated Ca II H lines, photometric data, and spectral line profile variations.

\subsection{Photometric variations} \label{subsec:pho}

\begin{figure*}
 \includegraphics[width=17cm]{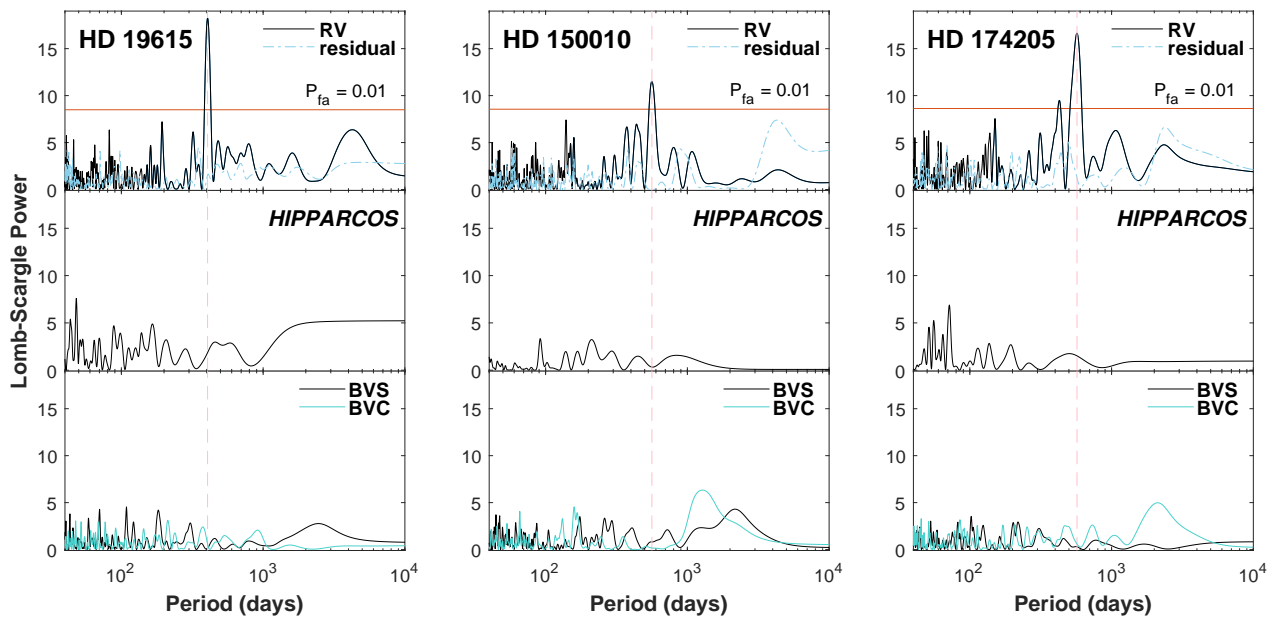}
\caption{L-S periodograms of the RV measurements and residuals (\emph{top}), \emph{HIPPARCOS} photometry (\emph{middle}), and the line bisectors of the span and curvature (\emph{bottom}) for the three stars. The vertical dashed  lines indicate the periods of each star.}
 \label{fig:ls}
  \end{figure*}
 
 \begin{figure}
   \centering
   \includegraphics[width=6cm]{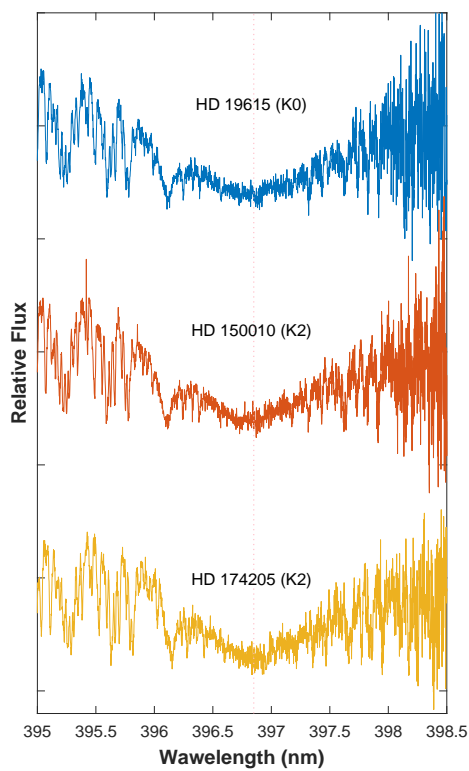}
      \caption{Spectra in the region of the Ca II H line. The vertical dotted line marks the center of the Ca II H profiles (\mbox{3968.5 \AA}).}
        \label{fig:ca}
   \end{figure}

The photometric data of \emph{HIPPARCOS} for our three target stars were obtained between November 1989 and March 1993. They are not contemporaneous with more recent BOES RV measurements. However, because photometric data cover the time interval relevant to a maximum period of 1071 days in our study, an examination of the \emph{HIPPARCOS} photometry may reveal periodic variations, possibly caused by rotational modulation of surface features or stellar oscillations. We thus used a total of 160, 143, and 119 observations of \mbox{HD 19615}, \mbox{HD 150010}, and \mbox{HD 174205}, respectively.

The rms scatters are 0.011, 0.007, and 0.011 mag for the three stars, respectively, which were comparable to the photometric uncertainties of \emph{HIPPARCOS} data. We also analyzed the L-S periodogram with photometric data and could not find any periodicity corresponding to the RV variations (middle panels in Fig.~\ref{fig:ls}).


\subsection{Line bisector variations} \label{subsec:bis}

An examination of spectral line shapes has become an important tool in identifying the origin of RV variations \citep{1998ASPC..154..311H}. Whereas the orbital motion by a companion causes a Doppler shift without a change in the line shape, rotational modulation of stellar surface structures gives rise to RV variations accompanied by line shape variations. We thus investigated the change in the shapes of spectral lines using two bisectors: the bisector velocity span (BVS), the difference in velocity between the 0.8 and 0.4 flux levels of the continuum value, and the bisector velocity curvature (BVC), the difference in the velocity span of the upper and lower half of the bisector.

In order to avoid contaminations from other spectral lines, I$_{2}$ absorption lines, and telluric absorption lines, we selected the lines in the spectral region 6000{\AA} -- 6800{\AA}. The selected lines, such as Fe, Mn, Ti, Ni, and Cr lines, were relatively deep under the 0.3 intensity level of the continuum value. We chose different lines for each star depending on the spectral type. Finally, we obtained the average bisector value by combining individual bisector measurements.
The L-S periodograms of the bisectors, both BVS and BVC, do not show any significant peaks associated with those of the RV variations (bottom panels in Fig~\ref{fig:ls}).

\subsection{Pulsations} \label{subsec:pul}

All target stars are evolved stars that are classified as K0, K2, or K2. Because most K giant stars have intrinsic mechanisms that may produce periodic RV variations, possible mechanisms such as pulsations should be investigated carefully \citep{2008A&A...480..215H}. Some K giant stars are known to be pulsating stars, which usually pulsate on several modes. Taking into account the K0-M3 classification and surface gravity of our stars, we can exclude known types of pulsations as the origin of the observed RV variations. The reason is that RV detected periods are much longer than the expected fundamental radial mode pulsations. The non-radial modes are more difficult to identify than the radial modes. We thus have to wait for a more detailed understanding of the new forms of oscillations that can be rotated to very long period oscillations (on the order of several years).

\subsection{Surface activity} \label{subsec:sur}
 A rotating star with surface features such as spots caused by chromospheric activity may exhibit periodic RV variations, which can be misinterpreted as a planetary signal \citep{2001A&A...379..279Q}. The Ca II H line (3968.5 $\AA$) has traditionally been used as an indicator of stellar activity \citep{1913ApJ....38..292E}. If there is chromospheric activity, an emission line will be shown at the center of the Ca II H absorption line profile. Several evolutionary stages of chromospheric activity surveyed recently around late-type evolved stars (\citealt{2005ESASP.560..963S,2008AJ....135.1117M,2009AJ....137.4282M}) have allowed us to investigate the oscillation of the magnetic activity of late-type giant stars.
We plotted the Ca II H line of each star in Fig.~\ref{fig:ca}. There are no noteworthy emissions at the core of Ca II H lines for any of the target stars.
Considering all the facts, we conclude that the regular RV variations in our target stars are most likely caused by orbiting planetary companions.

\section{Discussion}

We have detected long-period RV variations in three giant stars. The stars show no variations in line shape as measured by the spectral line bisectors.
Our target stars also seem to show a lack of variations in the \emph{HIPPARCOS} photometry. However, the lack of photometric variations
is only suggestive as the \emph{HIPPARCOS} measurements were not contemporaneous to our data.

Although the exact nature of long-secondary periods (LSPs) in giants is still unknown, \citet{2004ApJ...604..800W} speculated that the origin of long-period RV variations for LSP giants can be explained as a non-radial pulsation in a low-degree g$^{+}$ mode with star spot activities.
\citet{2015MNRAS.452.3863S} also suggested that the length of the LSPs was roughly consistent with the periods predicted from the non-radial pulsation in the oscillatory convective mode, which is a g$^{-}$ mode in a strongly non-adiabatic condition. However, the photometric variation for the stars may not be related to this mechanism, which operates in LSP giants.  Also, the amplitudes of RV variations are much smaller than those of typical LSPs.

\citet{2018AJ....155..120H} and \citet{2019A&A...625A..22R} both found RV signals that masqueraded as planetary companions with periods of a few hundred days and masses of a few Jovian masses around the K5 giants Gamma Draconis and Aldebaran. They concluded that the most likely cause seems to be a type of stellar oscillation in evolved K giants. Thus, further investigation into the origin of giant stars is needed before conclusions on the true cause of the RV variations in giant stars can be made.

On the other hand, \citet{2021ApJS..256...10D} showed that planet discoveries around stars with radii larger than 21 $R_{\sun}$ seem to
be rather problematic (this applies to HD~19615 and HD~174205, while HD~150010 is slightly below this radius limit).
Many of the stars with questionable planets have [Fe/H] < 0, which applies to all three stars of the current sample.
Finally, all these spurious planets have periods in the range between 300 and 800 days, which again applies to all three stars.
Stars with the given properties seem to show intrinsic RV variations of a so far unknown mechanism, which makes them very problematic targets for a planet search via the Doppler method.

Despite the possible verification of the cause of the long-term RV variations in giant stars at this point, the result of our research may be somewhat negative in light of recent research \citep{2021ApJS..256...10D}. Therefore, it is necessary to develop a new, improved interpretation method in the future.

Based on measurements conducted by \citet{2019A&A...623A..72K} of the proper motion anomaly between HIPPARCOS and Gaia DR2 main astrometry, the astrometric detection limits at 1 AU from this paper for HD~16915, HD~150010, and HD~174205 are $\sim$5$M_{J}$, $\sim$3.2$M_{J}$, and $\sim$11.4$M_{J}$, respectively. Of course, these are slightly below or above the reported RV minimal masses of the companions. The future Gaia DR4 astrometry will be even more detailed and precise.


\begin{acknowledgements}
BCL acknowledges partial support by the KASI (Korea Astronomy and Space Science Institute) grant
2021-1-830-08 and acknowledge support by the National Research Foundation of Korea(NRF) grant funded by the Korea government(MSIT) (No.2021283200).
M.G.P. was supported by the Basic Science Research Program through the National Research Foundation of Korea (NRF) funded by the Ministry of Education (2019R1I1A3A02062242) and KASI under the R\&D program supervised by the Ministry of Science, ICT and Future Planning.
 This research made use of the SIMBAD database, operated at the CDS, Strasbourg, France.
\end{acknowledgements}



\end{document}